%% file: elsarticle-template.tex
\documentclass[]{elsarticle}

\usepackage[margin=1.0in]{geometry}
\usepackage{array}
\usepackage{subcaption}
\usepackage{booktabs}
\usepackage{multirow}
\usepackage{tabularx}

\usepackage{ragged2e}
\newcolumntype{Y}{>{\centering\arraybackslash}X}

\usepackage{graphicx}
\usepackage{float}
\usepackage{amssymb}
\usepackage{fontawesome}

\usepackage{amsmath,bm}
\usepackage{relsize} 

\usepackage{hyperref}
\usepackage[usenames,dvipsnames]{xcolor}
\hypersetup{
    colorlinks,
    linkcolor={red!50!black},
    citecolor={blue!50!black},
    urlcolor={blue!80!black}
}

\usepackage{amsmath}
\usepackage{tikz}
\usetikzlibrary{automata}       
\usetikzlibrary{calc}           
\usetikzlibrary{fit}            
\usetikzlibrary{matrix}         
\usetikzlibrary{arrows.meta}    
\usetikzlibrary{shapes.geometric, arrows}
\usetikzlibrary{patterns}
\usetikzlibrary{backgrounds}

\usepackage{physics}
\usepackage[outline]{contour} 
\usetikzlibrary{angles,quotes} 
\contourlength{1.2pt}

\tikzset{>=latex} 
\usepackage{xcolor}
\colorlet{veccol}{green!70!black}
\colorlet{vcol}{green!70!black}
\colorlet{xcol}{blue!85!black}
\colorlet{projcol}{xcol!60}
\colorlet{unitcol}{xcol!60!black!85}
\colorlet{myblue}{blue!70!black}
\colorlet{myred}{red!90!black}
\colorlet{mygray}{black!5!white}
\colorlet{mypurple}{blue!50!red!80!black!80}
\tikzstyle{vector}=[->,very thick,xcol]
\tikzstyle{vector2}=[->,thick,myred]

\usepackage{pgfplots}
\usepackage{pgfplotstable}
    \pgfplotsset{
        compat=1.9,
        compat/bar nodes=1.8,
    }
\usepgfplotslibrary{statistics}
\tikzset{every picture/.style={/utils/exec={\ttfamily}}}

\tikzstyle{startstop} = [rectangle, minimum width=1.75cm, minimum height=1cm, text centered, draw=black, fill=gray!10]
\tikzstyle{process} = [rectangle, minimum width=1.75cm, minimum height=1cm, text centered, draw=black, fill=white]
\tikzstyle{arrow} = [thick,->,>=stealth]

\usepackage{color, colortbl}
\definecolor{Gray}{gray}{0.925}

\usepackage{listings} 
\usepackage{xcolor}
\definecolor{codegreen}{rgb}{0,0.6,0}
\definecolor{codegray}{rgb}{0.5,0.5,0.5}
\definecolor{codepurple}{rgb}{0.58,0,0.82}
\definecolor{backcolour}{rgb}{0.95,0.95,0.92}
\definecolor{light-gray}{gray}{0.95} 

\lstdefinestyle{mystyle}{
    basicstyle=\scriptsize\ttfamily,       
    backgroundcolor=\color{light-gray},   
    stepnumber=1,                   
    numbers=left,
    numbersep=5pt,                  
    backgroundcolor=\color{light-gray},  
    showspaces=false,               
    showstringspaces=false,         
    showtabs=false,                 
    frame=single,                   
    tabsize=2,                      
    literate={\ \ }{{\ }}1,         
    captionpos=b,                   
    breaklines=true,                
    breakatwhitespace=false,        
    xleftmargin=0.5em,              
    breakindent=0pt,
    breakatwhitespace,
    columns=fullflexible,
}

\makeatletter
\def\ps@pprintTitle{%
  \let\@oddhead\@empty
  \let\@evenhead\@empty
  \let\@oddfoot\@empty
  \let\@evenfoot\@oddfoot
}
\makeatother

\journal{Elsevier}

\begin{document}

\begin{frontmatter}

\title{Vector database management systems:\\Fundamental concepts, use-cases, and current challenges}

\author{Toni Taipalus}

\fntext[myfootnote]{To cite this article, please refer to the peer-reviewed, published version in Cognitive Systems Research \url{https://doi.org/10.1016/j.cogsys.2024.101216}}

\begin{abstract}
Vector database management systems have emerged as an important component in modern data management, driven by the growing importance for the need to computationally describe rich data such as texts, images and video in various domains such as recommender systems, similarity search, and chatbots. These data descriptions are captured as numerical vectors that are computationally inexpensive to store and compare. However, the unique characteristics of vectorized data, including high dimensionality and sparsity, demand specialized solutions for efficient storage, retrieval, and processing. This narrative literature review provides an accessible introduction to the fundamental concepts, use-cases, and current challenges associated with vector database management systems, offering an overview for researchers and practitioners seeking to facilitate effective vector data management.
\end{abstract}

\begin{keyword}
vector \sep database \sep feature \sep challenge \sep neural network \sep deep learning
\end{keyword}


\end{frontmatter}

\section{Introduction}
\label{sec-introduction}

It is increasingly common that rich, unstructured data such as large texts, images and video are not only stored, but given semantics through a process called \textit{vectorization} \cite{Wang_2021} which captures the features of the data object in a cost-effectively processed numerical vector such as $\vec{k} = [6,7]$. The vectors are \textit{n}-dimensional, and consist of natural, real, or complex numbers, where one number represents a feature or a part of a feature. The features that form a vector can range from simple, such as the number of actors in a stage play, to complex, such as textures identified in an image by a neural network \cite{Gasser_2020}, where number 3 may correspond to texture of human skin, while number 10 may correspond to the texture of a cat's fur. In contrast to traditional data models such as relational, where queries often take forms such as ``\textit{find the orders of a specific user}'' or ``\textit{find the products that are on sale}'', vector queries typically search for \textit{similar} vectors using one or several query vectors. That is, queries take forms such as ``\textit{find ten most similar images of cats that look like the cat in this image}'' or ``\textit{find the most suitable restaurants for me given my current position}''. 

Managing vector data has gained increased popularity, partly due to applications such as reverse image search, recommender systems, and chatbots, and this trend is on the rise \cite{Li_2023}. Consequently, efficient management of data requires a dedicated database management system (DBMS). A vector DBMS (VDBMS) is not strictly a requirement for any business domain, as vectors can be stored and queried without a dedicated DBMS, similarly to relational or document data can be stored and queried without a relational DBMS. The DBMS, however, in all cases, facilitates data management that is \textit{feasible}, freeing development resources towards other business domain critical tasks by providing ready-made features such as transaction and access control, automated database scalability, and query optimization. Additionally, increasingly complex business domains require increasingly complex features such as vector similarity search complemented by metadata filters, as well as searching with multiple query vectors \cite{Wang_2021}, and efficient ways to manage access control and concurrent transactions.

This narrative literature review aims to provide an easily accessible description of fundamental concepts behind VDBMSs (Section~\ref{sec-vectors_main}) without focusing on the intricacies of a single product, an overview of current VDBMS products and their features (Section~\ref{sec-products}), explanations behind some popular use-cases such as image similarity search and long-term memory for chatbots (Section~\ref{sec-use-cases}), and some of the current challenges related to VDBMSs (Section~\ref{sec-challenges}). This work assumes that the reader is familiar with fundamentals of some other type of database management system (e.g., relational), and does not detail the mathematics of vectors, or algorithms behind vector search or vector index creation.


\section{Vectors and vector database management systems}
\label{sec-vectors_main}

\subsection{Vectors as data representations}
\label{sec-vectors}

\begin{figure*}[t]
    \centering
    \begin{subfigure}[t]{0.45\textwidth}  
        \centering 
        \input{figs/fig-coordinates1}
        \vskip\baselineskip 
        \caption{Closeness in geospatial locations}
        \label{fig-coord1}
    \end{subfigure}
    ~
    \begin{subfigure}[t]{0.45\textwidth}  
        \centering 
        \input{figs/fig-coordinates2}
        \vskip\baselineskip 
        \caption{Closeness in genres}
        \label{fig-coord2}
    \end{subfigure}
    \caption{Simple examples of applications of two-dimensional vectors}
    \label{fig-vectors}
\end{figure*}
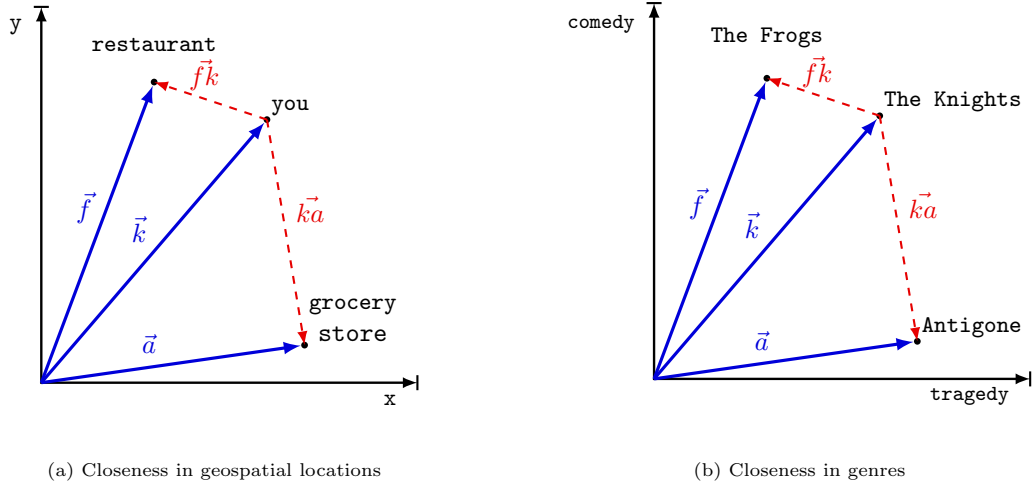

Perhaps one of the most intuitive use-case for vector data is in geospatial applications \cite[e.g.,][]{Touya_2020}. Two-dimensional points such as the location of the end-user and points-of-interest may be represented as vectors, and the closest points-of-interest may be calculated with simple and well-understood operations. For example, by calculating the distance between the end-user (``\textit{you}'' in Fig.~\ref{fig-coord1}) and points-of-interest, the length of vectors ($\vec{fk}$ and $\vec{ka}$, i.e., distance) can be compared, and the closest point-of-interest found. If we consider the vector for the end-user as $\vec{k} = [6,7]$, the vector for the restaurant as $\vec{f} = [3,8]$ and the vector for the grocery store as $\vec{a} = [7,1]$, we can calculate the similarity or closeness of the vectors by, e.g., Euclidean distance or cosine similarity.


In addition to coordinates, other types of data can be represented as vectors. For example, instead of coordinates, Fig.~\ref{fig-coord2} shows Greek plays mapped along how comic and tragic they are. By examining closeness based on these two dimensions, we can, e.g., calculate that the play \textit{The Knights} is closer to \textit{The Frogs} than \textit{Antigone}, that is, the vector $\vec{fk}$ is shorter than the vector $\vec{ka}$. The vector for \textit{Antigone} can be represented as $\vec{a} = [7,1]$, where the first component represents the amount of tragedy, and the second component the amount of comedy. Closeness of the vectors is not the only way to measure similarity.

The aforementioned are examples of very low-dimensional vectors. By increasing the dimensions of vectors (say, by adding \textit{z} coordinates, or another genre, \textit{drama}), vectors can capture increasingly rich data. If \textit{Antigone}'s drama amounts to 6, the vector for \textit{Antigone} in three-dimensional space is $\vec{a} = [7,1,6]$. Furthermore, a high-dimensional vector may have thousands or millions of dimensions, making the visualizations of such vectors unfeasible, and the data unreadable for a human. Such high-dimensional vectors can be used to represent more complex data such as text, image, audio and video features. From data-representation perspective, this separates vector databases from relational and NoSQL databases, in which data objects are often human-readable, contextualized numbers, text strings, and time. This holds especially in relational databases, where data objects are given meaning by table and column names. In NoSQL databases, data objects may also be highly unstructured and more difficult to understand for a human.

\subsection{Vector database management systems}
\label{sec-vdbmss}

A vector database management system is \textit{a specialized type of database management system that focuses primarily on the efficient management of high-dimensional vector data}. Similarly to other types of database management systems (such as relational, document, and graph), this definition requires that a VDBMS is functional software that can manage data, rather than being merely, e.g., a software library. Data management includes but is not limited to data querying and manipulation, collection of metadata, indexing, access control, backups, support for scalability, and interfaces with other systems such as database drivers, programming languages, frameworks, and operating systems. Furthermore, a VDBMS focuses on the management of vector data. There are several DBMSs that offer support for multiple data models (e.g., PostgreSQL supports relational, document and object-oriented data models\footnote{https://www.postgresql.org/docs/16/ddl-inherit.html}, and Redis supports key-value and vector data models\footnote{https://redis.io/docs/get-started/vector-database/}), yet the primary focus of such systems is typically on one data model. It has been noted that such systems miss optimization opportunities for vector data, and may lack features such as the use multiple query vectors \cite{Wang_2021}. Finally, a VDBMS focuses on the management of high-dimensional vectors. Systems focusing on, e.g., two or three-dimensional geospatial data management are not considered VDBMSs in this context.

VDBMSs typically support similarity search through indexing methods that enable rapid and accurate searching of similar vectors, i.e., search for vectors that closely resemble a given query vector based on specific distance metrics such as Euclidean distance or cosine similarity. This capability is particularly valuable in various applications where finding similar vectors is crucial, such as image or text retrieval systems. VDBMSs also offer support for vector operations, allowing users to perform mathematical computations on vectors. These operations may include arithmetic calculations, statistical analysis, or transformations to manipulate the vectors. In colloquial language, the term \textit{vector database} is sometimes used as a synonym for a VDBMS despite the fact that a VDBMS is software, yet a vector database is a collection of data. It is also worth noting that despite their popularity, we -- among others \cite{Wang_2021} -- do not consider algorithms or libraries such as Facebook's FAISS library \cite{Johnson_2021} VDBMSs, as they do not provide many of the functionalities described above.

\subsection{Database system architecture}
\label{sec-architecture}

\begin{figure}[t]
    \centering
    \input{figs/fig-dbs}
    \caption{A simplified view of a database system illustrating the flow and transformation of information to and from the vector database; the vectorization process transforms information into vectors which can be quickly compared with each other; it is worth noting that the natural language query depicted here requires data additional to the actual plays}
    \label{fig-dbs}
\end{figure}
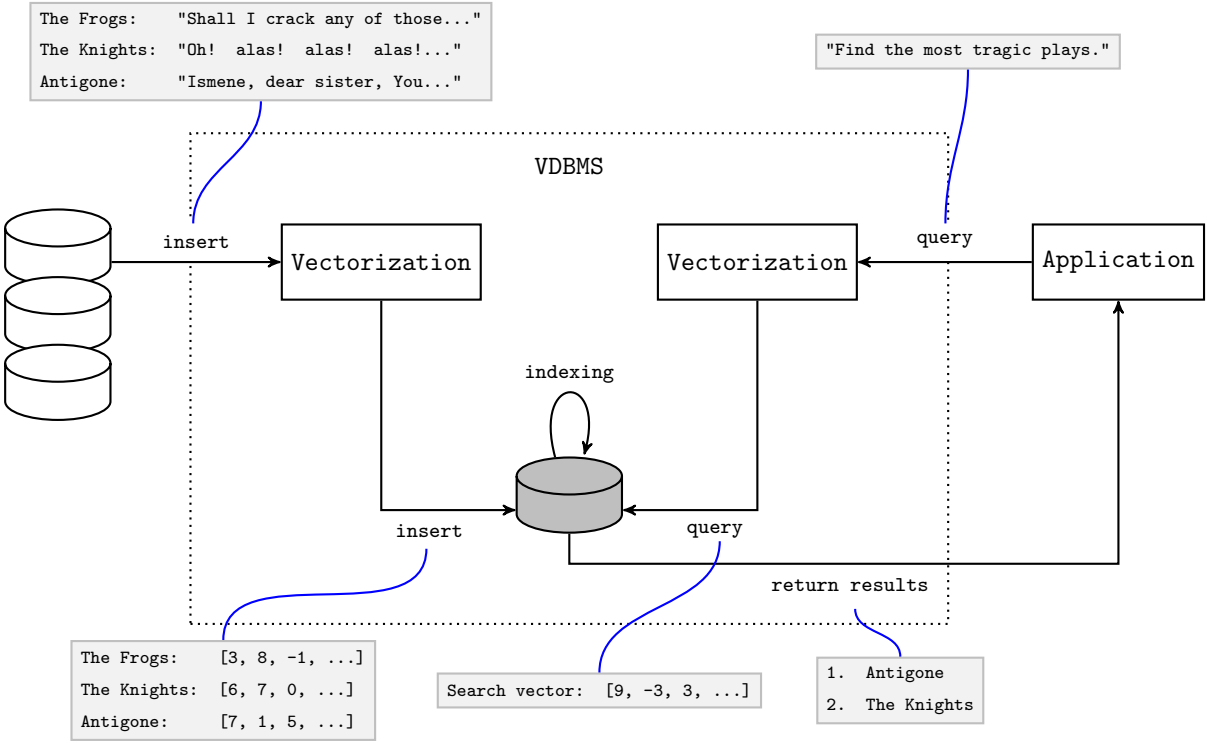

A database system consists of one or several database management systems, databases, and software applications. Fig.~\ref{fig-dbs} shows a simplified flow of information from traditional data sources (depicted on the left-hand side, e.g., relational databases) to the vector database (gray). Continuing with the example of Greek plays, the human-readable texts of the plays are \textit{vectorized}, i.e., transformed into high-dimensional vector representations in a way that captures meaningful relationships or patterns. The outcome of the vectorization process, i.e., the vector, is often called a \textit{vector embedding} or a \textit{feature vector}. In addition to the vector itself, VDBMSs typically store a vector identifier, some vector metadata, and possibly the data that the vector represents. For example, for the play \textit{Antigone}, a VDBMS may store an unique identifier, a vector embedding of the text of the play that contains numerical data about the amount of tragedy, comedy and drama in \textit{Antigone}, metadata such as type of work (``\textit{play}'') and country of origin (``\textit{Greece}''), and the play itself in plain text. The play itself is often referred to as the \textit{payload} of the vector. As another example, YouTube utilizes metadata such as user and video language and time since video was last watched in providing personalized recommendations \cite{Covington_2016}.

In natural language processing, words and phrases are vectorized into vectors in such a way that similar words have similar vector representations. Similarity can mean different things depending on the context, e.g., words may sound similar (\textit{walking} and \textit{talking}), or words may mean similar things (\textit{walking} and \textit{running}) in different contexts. The amount of tragedy in a play may depend on the number of tragic words in the play, a sentiment analysis assessing the tone of the play, or topic modeling which identifies key themes in the play. This process helps algorithms understand and work with the data more effectively. \textit{Word2vec} \cite{Mikolov_2013}, \textit{FastText}, and \textit{Doc2vec} \cite{Le_2014} are examples of techniques that create vector embeddings for words in natural language. 

\begin{table}[t]
\centering
\begin{tabularx}{\textwidth}{XYYYY}
\hline
\hline
 & Characteristics & Use-cases & Advantages & Disadvantages \\
\hline
\hline
Product Quantization & Divides vectors into smaller parts & Image search & Reduces dimensionality & Lossy compression may reduce accuracy \\
\hline
Locality-Sensitive Hashing & Hashes similar vectors to same buckets & Near-duplicate detection & Enables approximate similarity search & Requires parameter tuning \\
\hline
Hierarchical Navigable Small World & Creates a hierarchical graph & Recommendation systems, text search & Fast neighborhood exploration & Complex index structure, space overhead \\
\hline
R-trees & Hierarchical structure with bounding boxes & Spatial data (geospatial indexing) & Efficient range queries, updates & Slower nearest-neighbor searches \\
\hline
KD-trees & Binary tree partitioning along dimensions & Machine learning, clustering & Balanced tree structure, good for low dimensions & Inefficient in high dimensions, complex build \\
\hline
Random Projection & Projects high-dimensional data randomly & Text classification, clustering & Fast indexing, good for high dimensions & May lose information, requires tuning \\
\hline
\hline
\end{tabularx}
\caption{Comparison of some vector indexing techniques}
\label{tab:vector-indexing-comparison}
\end{table}

After the data objects have been vectorized and stored in the vector database, the data are indexed to enable faster queries, as with effectively all data models \cite{Kraska_2018}. As vector queries are almost always approximations, one of the primary trade-offs between different indexing algorithms are accuracy and speed. Some popular algorithms are \textit{Product Quantization} \cite[e.g.,][]{Ge_2013,Jegou_2011}, which divides high-dimensional vectors into smaller parts and summarizes each part separately, reducing dimensionality and storage space requirements, but losing some accuracy, \textit{Locality-Sensitive Hashing} \cite[e.g.,][]{Zheng_2020}, which hashes similar vectors to the same buckets, enabling approximate similarity search, and \textit{Hierarchical Navigable Small World} \cite[e.g.,][]{Zhao_2020,Malkov_2020}, which creates a hierarchical graph with fast neighborhood exploration by building a small world network. Other algorithms include \textit{R-trees} \cite{Guttman_1984}, \textit{KD-trees} \cite[e.g.,][]{Silpa_Anan_2008}, and \textit{Random Projection} \cite[e.g.,][]{Dasgupta_2008}. Table~\ref{tab:vector-indexing-comparison} summarizes various vector index types. It is worth noting that the choice of indexing algorithm depends on the data characteristics, dimensionality, and search requirements. Index creation is typically computationally expensive.

Similarly to inserting data into the vector database, queries in natural language or human-readable values in computer language queries must be vectorized before the VDBMS can assess vector similarity. The vectorization may happen in the application program or the VDBMS (the latter case is depicted in Fig.~\ref{fig-dbs}, yet the former case is more typical). Vectorization can be done in multiple ways depending on the data and the purpose of the vectorization. Despite the fact that feature vectors of Greek plays and feature vectors of images of cats may look similar (i.e., both are ``lists'' of numbers), their values represent different things.

As mentioned earlier, vector similarity or closeness may be assessed using several methods such as \textit{Jaccard similarity}, which measures the similarity between two sets (i.e., vectors) by comparing their shared elements to the total number of distinct elements in both sets, \textit{Euclidean distance} (L2), which measures the straight-line distance between two points in a space with multiple dimensions, \textit{dot product}, which computes the sum of the products of corresponding elements in two vectors, or \textit{cosine similarity}, which measures the cosine of the angle between two vectors, indicating how similar their directions are regardless of their magnitudes. The choice of method depends on the context and the specific characteristics of the data. For more in-depth, mathematical explanations, Wang \textit{et al.} \cite{Wang_2016} provide accessible overview of several indexing and search methods mentioned above.

From a developer perspective, queries in VDBMSs are more closely related to simple document or key-value store queries than to complex queries in relational databases. Instead of retrieving documents based on document identifiers as in many NoSQL systems, vectors are retrieved using one or several query vectors. Despite this similarity in queries, the query execution internals differ, since VDBMS queries typically search for nearest neighbor vectors instead of exact matches. Fig.~\ref{fig-queries} illustrates some basic queries in three VDBMSs and in PostgreSQL with the \textit{pgvector} extension. Instead of searching for Greek plays where the amount of tragedy is high, as one probably would with a relational query, a vector query may retrieve Greek plays which are similar to a particular play in terms of tragedy, comedy, drama, author, publication year, etc.

\begin{figure*}[t]
    \input{figs/fig-queries}
    \caption{Hybrid queries in different VDBMSs using Python, and in PostgreSQL using SQL}
    \label{fig-queries}%
\end{figure*}

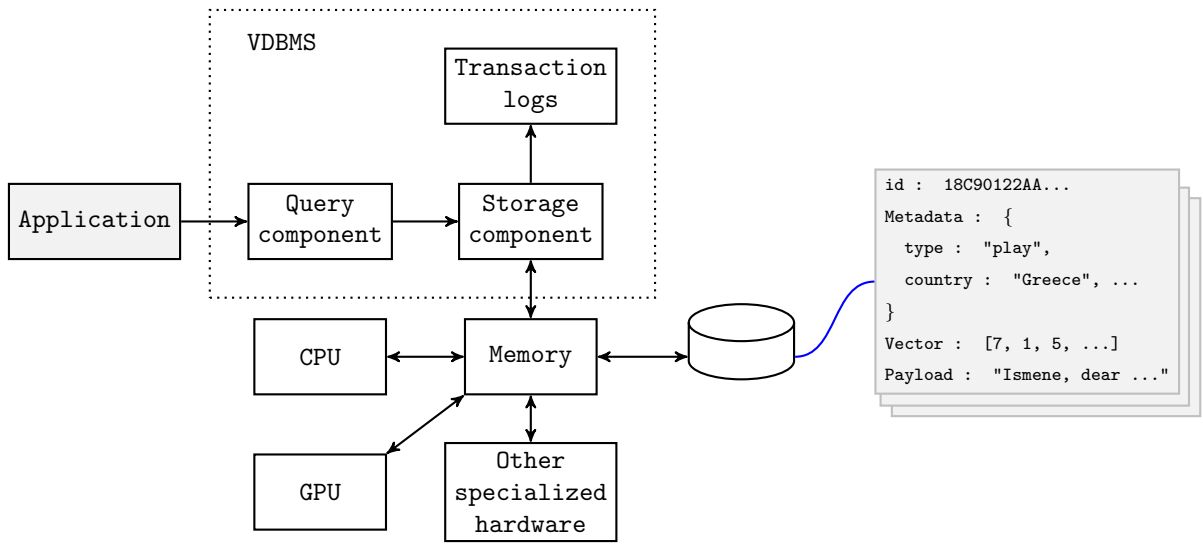
\begin{figure}[t]
    \centering
    \input{figs/fig-dbms}
    \caption{A generalized overview of VDBMS components; the arrows represent the flow of information from the software application through the VDBMS to the physical database; the database represents persistent storage device, contrary to Fig.~\ref{fig-dbs}, where the database represents the logical database structure maintained by the VDBMS; the righ-hand side shows an example of the stored data object consisting of metadata, the vector, and vector payload}
    \label{fig-dbms}
\end{figure}

In addition to the vectors themselves, queries may utilize metadata to, e.g., limit the number of vectors to compare. For example, if the end-user is requesting data on Greek plays, and the database contains metadata for language and type of media, the vector similarity search may be limited to Greek plays rather than all written art originating from all countries. While vectors are indexed using different vector indices, metadata may be indexed using more traditional techniques such as \textit{B$^+$-trees} to support range queries. Queries that utilize both a query vector and metadata filters are called \textit{hybrid queries}. If a VDBMS does not provide the means for hybrid queries, metadata-based searches may be implemented separately as part of a broader architecture. Fig.~\ref{fig-dbms} provides a generalized (i.e., not product-specific) overview of VDBMS components. These components are in principle similar to components in other types of DBMSs: the query component parses the queries and other statements from the software application, checks user access rights on data object level, optimizes the query, and passes the query to the storage component. The storage component logs the transaction if the VDBMS utilizes transaction logs such as Write Ahead Logging, manages transaction locks if applicable, and retrieves or stores the data the application has requested with the help of different buffers, memory, CPU and GPU, and possibly other specialized hardware such as Field-Programmable Gate Arrays or tensor processing units.

\section{Products and features}
\label{sec-products}

At the time of writing, DB-Engines \cite{db-engines} lists seven VDBMSs: Pinecone, Chroma, Milvus, Weaviate, Vald, Qdrant and Deep Lake. However, since Vald primarily focuses on similarity search and lacks features such as access control and integrations to other technologies, we considered Vald a vector search engine rather than a VDBMSs as defined in Section~\ref{sec-vectors}. Several of these products are designed from the ground up to utilize different types of processing units or devices, and multi-GPU and CPU parallelism in a coordinated manner \cite{Wang_2021}. The VDBMSs typically implement several index and search methods (such as Euclidean distance), and the optimizer component selects the most suitable search method depending on the characteristics of the data and the query, similarly to the optimizer in relational DBMSs.

In addition to the VDBMSs mentioned above, there are also several DBMSs with multiple data models, vectors being one of them, several vector extensions to other DBMSs such as PostgreSQL, MongoDB, Cassandra, Redis and SingleStore, and as vector database-enabling libraries for programming languages, such as Thistle for Rust \cite{Windsor_2023}. Table~\ref{tab-features} lists some features of these six VDBMSs. Similarly to NoSQL systems, we expect VDBMSs to develop rapidly in terms of features, new products, and community support.

\begin{table}[t]
\centering
\footnotesize
\caption{VDBMS features; example use-cases are based on a product's documentation's use-case examples as of August 2023}
\begin{tabularx}{\textwidth}{lccc}
\toprule
             & License                    & First release    & Querying with metadata     \\
\midrule
Pinecone     & Proprietary                & 2021             & rich expressions           \\
Chroma       & Apache 2.0                 & 2023             & rich expressions           \\
Milvus       & Apache 2.0                 & 2019             & rich expressions           \\
Weaviate     & BSD 3-clause / proprietary & 2019             & supported                  \\
Qdrant       & Apache 2.0 / proprietary   & 2022             & rich expressions           \\
Deep Lake    & Apache 2.0 / proprietary   & 2019             & rich expressions           \\
\midrule
             & Integration                & Querying         & Example use-cases          \\
\midrule
Pinecone     & OpenAI, LangChain, others   & Java, Python, C\#, several others     & chatbots, image search \\
Chroma       & LangChain, LlamaIndex       & JavaScript, Python, Ruby, others      & chatbots               \\
Milvus       & OpenAI, LangChain, others   & Java, Python, Go, Node.js             & chatbots, image/audio/video search  \\
Weaviate     & OpenAI, Cohere, PaLM        & Java, JavaScript, Python, Go, GraphQL & chatbots, image search \\
Qdrant       & OpenAI, LangChain, others   & Python, JavaScript, Go, Rust          & chatbots, image search \\
Deep Lake    & LlamaIndex, LangChain       & Python, SQL-like TQL                  & image search           \\
\bottomrule
\end{tabularx}
\label{tab-features}
\end{table}

\section{Use-cases}
\label{sec-use-cases}

\subsection{Similarity search in general}
\label{sec-uc-general}

As explained in Section~\ref{sec-vectors}, there are many use-cases for vector data. Effectively all data objects that can be vectorized in a meaningful way may be used in approximate similarity search, which is the basis for almost all vector database retrieval operations. Although the next subsections focus on some popular use-cases for vector databases, it is worth noting that this is not an exhaustive list. For example, vectors are used in storing and comparing molecular structures \cite{Mater_2019} and rentable apartments \cite{Grbovic_2018}, automated black-and-white image colorization \cite{Baldassarre_2017}, facial expression recognition \cite{Bashyal_2008}, tracking digital image assets \cite{Sahoo_2023}, and recommender systems \cite{Shankar_2017}.

\subsection{Image and video similarity search}
\label{sec-uc-img-vid}

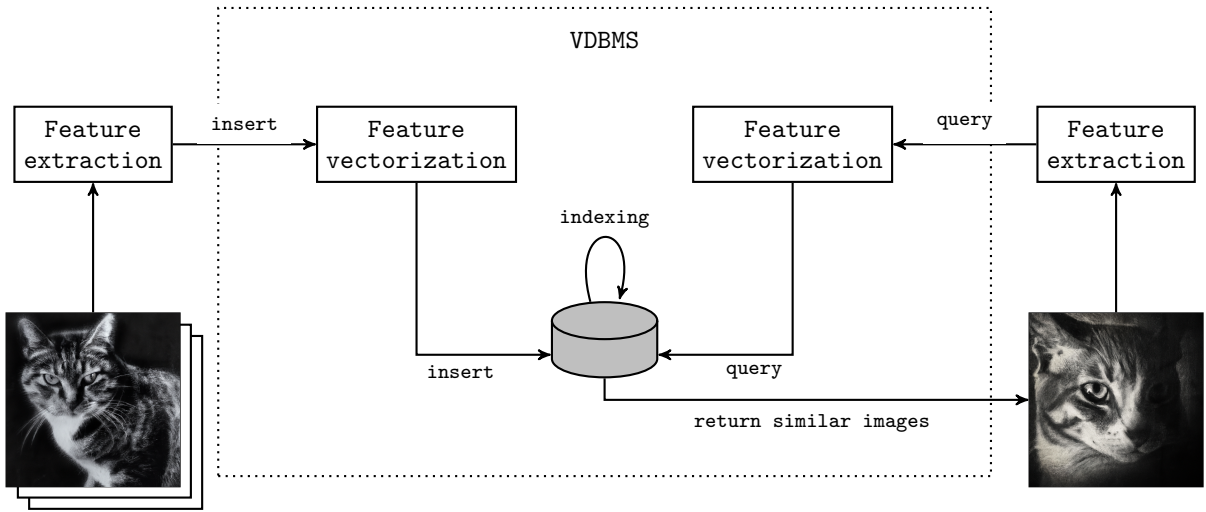
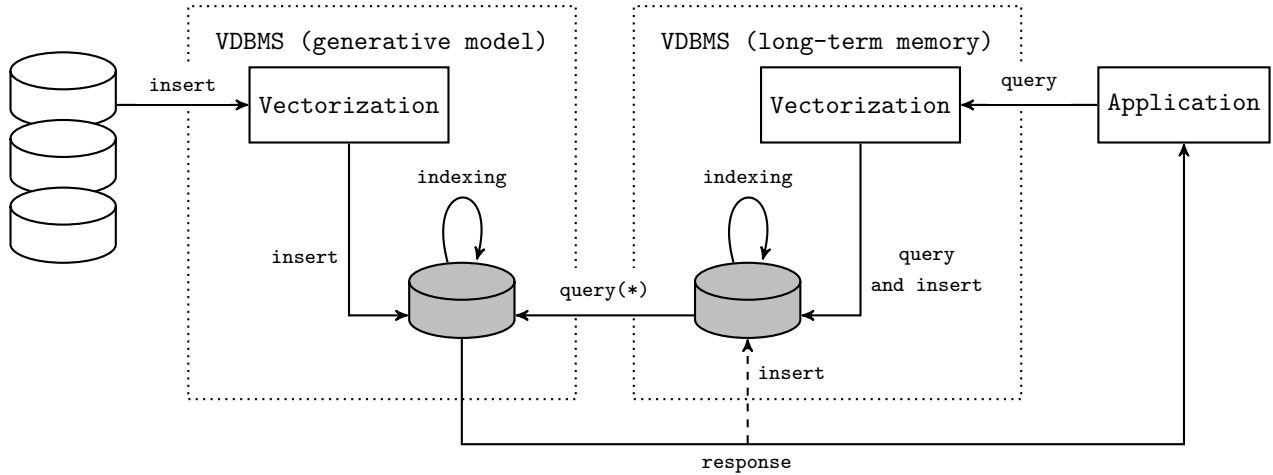
\begin{figure}[t!]
\centering
\subfloat[By vectorizing images (left-hand side) by their features, reverse image search can be used to find similar images (right-hand side)\label{fig-use-cases-img}]{%
  \input{figs/fig-u-img} }
\vspace*{5mm}
\subfloat[A VDBMS can be used for long-term storage of chatbot conversations, which can be then queried and used as additional context for generative models; the query with an asterisk contains both the original user prompt as well as similar conversations, which are both used as query vectors for the generative model\label{fig-use-cases-chat}]{%
  \input{figs/fig-u-chat} }
\caption{Uses-cases for VDBMSs in the domains of image similarity search and chatbots; note how here all the VDBMSs handle the vectorization of data -- this is not usually the case}
\label{fig-use-cases}
\end{figure}

In a similar fashion as Greek plays, images can be vectorized, yet the process is typically more complex and involves image normalization in terms of size and pixel values, and feature extraction prior to vectorization. Feature extraction, which is typically external to the VDBMS, can involve passing the images -- one at a time -- through a convolutional neural network. The process extracts increasingly abstract features from the image, starting from simple features such as the presence of vertical and horizontal edges and simple shapes \cite[e.g.,][]{Herbulot_2006}, to textures such as fur, foliage and water. These features are vectorized and used for similarity search. Images with similar vector representations are likely to be visually similar in terms of the captured features. Similarly to Fig.~\ref{fig-queries}, Fig.~\ref{fig-use-cases-img} shows that after a set of images has been vectorized, similar process can be used for reverse image search (i.e., searching for images with image input). Once similar vectors have been found, the VDBMS returns the vector payloads to the application.

In the context of video vectorization, videos are typically broken down into individual frames, although just a representative subset of frames may be considered. Similarly to stand-alone images, features are extracted from the individual frames and vectorized into a feature vector representing its content. Additionally, temporal information is often needed to further understand the contents of the video. For example, one video of a Greek play may develop from comedy to tragedy, while another may do the opposite. Without temporal information about the order of the frames, it is not possible to tell one from another in this regard. The process results in a sequence of feature vectors which can be combined. In other words, the sequence of vectors may be considered a three-dimensional tensor, where dimensions represent frames, features, and time. This tensor can be stored as a flattened vector in the vector database. It is worth noting that the \textit{dimension} of the tensor here is a different concept than the \textit{dimension} of the vector.

\subsection{Voice recognition}
\label{sec-uc-voice}

Voice recognition using vectors works in a similar fashion to video vectorization and search. If the audio is analog format, it is digitized and divided into short frames, each of which represent a segment of the audio. Each frame is normalized, filtered and transformed with various techniques, and finally stored as a feature vector \cite{Venya_1998}. The whole audio is therefore a sequence of feature vectors, all together representing a spoken word or sentence, a song, or some other type of audio. If voice recognition is used in user authentication, similar process may be applied to a spoken keyphrase, and the vectorized spoken keyphrase compared with vectorized recordings. On the other hand, if voice recognition is used in a conversational agent, the sequences of vectors can be used as an input for, e.g., neural networks to recognize and classify spoken words, and to respond accordingly in text or synthethized voice using a generative model such as ChatGPT. The examples of returning similar-looking cats and using voice recognition to authenticate users serve as opposing prime examples of tolerance in similarity search. While several images of similar cats may be returned with relatively low tolerance, authenticating an user by their voice requires high tolerance. 

\subsection{Chatbots and long-term memory}
\label{sec-uc-chat}

VDBMs can be used for long-term memory of chatbots or other generative models. Illustrated on the left-hand side of Fig.~\ref{fig-use-cases-chat}, a large dataset is first used to train a model capable of imitating understanding and producing natural language to a certain degree. A VDBMS can be used to store and index the vectors, although this can be done in other ways as well. Generative models have limitations when it comes to remembering past conversations or context, and currently, several technical limitations contribute to this challenge. For example, several models can only consider a limited amount of preceding text when generating a response. Consequently, they currently often struggle to recall detailed information from long conversations \cite{Tay_2020}. Generative models do not have a built-in memory of past interactions, and generate responses based on the immediate context provided in the input. Once a conversation becomes too lengthy or complex, the model's ability to reference earlier parts of the conversation diminishes. Furthermore, generative models are trained on large datasets, but they lack the ability to distinguish between factual information and user-specific interactions. This can lead to instances where the model provides inconsistent or incorrect information based on the training data \cite[cf. e.g.,][]{Zhang_2023}.

To counter these limitations, a VDBMS may be used as a long-term memory in such use-cases. As illustrated on the right-hand side of Fig.~\ref{fig-use-cases-chat}, when an end-user prompts (i.e., submits a query to) a chatbot, the natural language query is vectorized and used as a query vector for a long-term memory vector database to find top-\textit{k} similar conversations. The query (i.e., the user prompt in vectorized and natural language form) is also stored in the long-term memory vector database. Next, the original user prompt as well as \textit{k} similar past conversations are used as query vectors for the generative model (marked with an asterisk in Fig.~\ref{fig-use-cases-chat}). The generative model then generates a response, which is inserted in the long-term memory database in vectorized and natural language form, and returned to the application (i.e., the chatbot user interface). This approach not only allows the chatbot to remember past conversations, but also enables personalized information, conversation sequence encoding, and timestamps through vector metadata, and potentially reduces the use of computational resources without the need to retrain or fine-tune the generative model. In addition or alternatively to storing past conversations, a VDBMS can be used to store documents which are used as additional, context-providing input to the generative model. These documents can be private to the organization using the VDBMS, or they may be additional, timely information not included in the generative model. This approach is dubbed \textit{retrieval augmented generation} \cite{Cai_2022,Lewis_2020}.

In summary, the previous subsections illustrate different use-cases for VDBMSs, yet it can be seen that the function of the VDBMS is rather uniform regardless of the use-case. That is, from a transaction processing perspective, the VDBMS stores, indexes and retrieves vectors, and domain-specific processes such as image feature extraction are carried out in other parts of the system.

\section{Current challenges}
\label{sec-challenges}

\subsection{Balancing between speed and accuracy}
\label{sec-chall-balance}

As most queries in VDBMSs operate by searching approximate nearest neighbors, balancing between query response time and the accuracy of the results is a trade-off largely dictated by the business domain. Some vector index types such as \textit{Product Quantization} save storage space and speed up queries by abstracting and aggregating information with the cost of accuracy, while other index types such as \textit{R-trees} are lossless. Lossless indices are preferred when exact similarity measurements are critical, while lossy indices are used when approximate similarity searches are acceptable, and there is a need to reduce storage and computation costs.

The challenge in choosing between speed and accuracy is two-fold. First, compared to many other data models, the concept of query accuracy plays a significantly larger role. Although many NoSQL data models forsake data integrity for eventual consistency, the effects of such design principles for the end-user are relatively small compared to inaccurate vector searches. On the other hand, VDBMSs disregard many challenges related to other data models, such as the complexity of querying in relational databases, and respective challenges and complexities in logical database design in both relational and NoSQL data models. Second, the trade-offs between speed and accuracy are emphasized in especially large datasets where both speed and query accuracy are critical. For example, natural language operated decision support systems which use large corporate datasets or stock market data need to provide decisions fast, but without returning inaccurate or untrue results. One possible solution for ensuring both speed and accuracy is utilizing several indices for the same vectors, yet this approach naturally requires more storage capacity.

\subsection{Growing dimensionality and sparsity}
\label{sec-chall-dimen}

The growing dimensionality of vectors is a challenge. As the needs of the domain grow, it is natural that the vectorized data need more features. For example, it is reasonable to assume that a vector database of Greek plays will soon require more insights on the plays besides the amount of comedy and tragedy. This leads to increased storage requirements and computational complexity. 

Increased dimensionality also impacts similarity search, as the notion of proximity becomes less reliable in high-dimensional spaces. Euclidean distance, which is commonly used in low-dimensional spaces, becomes less reliable in high-dimensional spaces due to the concentration of points around the surface of the space. That is, in high-dimensional spaces, the volume of space grows exponentially with new dimensions, while the possible number of vectors typically does not. This results in vectors naturally concentrating close to the hypersurface of the space, as that is where the majority of the space is. Because vectors are concentrated near the surface, the distances between the vectors tend to be more similar in high-dimensional space \cite{Indyk_1998}. Developing effective distance measures that can capture the true similarity or dissimilarity between high-dimensional vectors is an ongoing research challenge.

Another challenge is the increased sparsity of high-dimensional vectors. As the number of dimensions grows, the available space becomes more sparsely populated, meaning that data points are spread out across the vector. For example, if a vector database consisting of feature vectors of images of cats is extended to cover images of other animals as well, vector dimensions associated with cats (or mammals, or chordates, etc.) do not contribute significantly to the overall structure of vectors depicting other animals. This sparsity complicates indexing and retrieval, as methods designed for denser data struggle to efficiently represent and query sparse data. The challenges associated with sparse data have been addressed in, e.g., the column-family data model, but not in the degree that is required with high-dimensional sparse vectors.

\subsection{Achieving general maturity}
\label{sec-chall-maturity}

DBMSs are typically large and complex pieces of software. It follows that there are several aspects to DBMSs that evolve and mature over time, and because VDBMSs are relatively novel systems (cf. Table~\ref{tab-features}), considerations such as their stability, reliability, and optimization are subject to even drastic development. In comparison, even mature relational DBMSs still receive critical bug fixes \cite{Oracle_bug_fixes_2023}.

Maturity is not a goal in on itself. Decades of development and testing have likely addressed many bugs and stability issues, making DBMSs in general more reliable for mission-critical applications. Over time, DBMSs tend to accumulate a rich set of features and functionalities. They often support a wide range of data models, query languages, and storage options, allowing them to cater to diverse use-cases. This is not necessarily the case with more novel VDBMSs. Additionally, a mature DBMS typically has a large and active user community, which can be valuable for getting support, finding resources such as online tutorials, and leveraging third-party extensions and integrations. 

Information security is a challenge that is not limited to business domains of VDBMSs. Due to their common use-cases, a vector database may contain sensitive information such as conversations intended to be private, biometric data, risk assessment profiles, and geospatial intelligence data. While many similar use-cases are common in relational databases as well, older DBMSs have had time to identify and address more security vulnerabilities, and usually have robust security features and practices in place. 

In summary, there are several open challenges regarding VDBMSs, some of which are related to algorithms such as the need for novel index structures, some to software such as the availability of certain features in VDBMSs, and some to social aspects such as the maturity and availability of online support. In the future, we expect the demand for vector databases to grow. Consequently, we expect VDBMS vendors to focus on developing and applying new algorithms for vector indices, as well as making high-dimensional vectors more human-readable through visualizations. Additionally, as data-intensive computational models are computationally expensive to retrain, we expect that VDBMS vendors will try to address this by implementing features for incremental learning, i.e., cost-effective fine-tuning of computational models.

\section{Conclusion}
\label{sec-concl}

Vector database is a growing data model intended for storing vectors which describe rich data in high-dimensional vectors. This study provided an overview of fundamental concepts behind vector databases and vector database management systems, such as different types of vector similarity comparison types, different vector index types, and the principal software components in a VDBMS. Additionally, this study described some VDBMSs and their features, as well as some popular use-cases for vector data such as chatbots and image similarity search. Finally, this study discussed some of the current challenges associated with VDBMSs such as high-dimensionality and sparsity of vector data, and the relative novelty of VDBMS products and the implications therein.

\section*{Acknowledgements}
Bitmap images of cats in Fig.~\ref{fig-use-cases} were generated with DALL-E.

\bibliography{sample-base}


\end{document}

%% file: figs/fig-coordinates1.tex
\begin{tikzpicture}
  \coordinate (O) at (0,0);
  \coordinate (R) at (1.5,4);
  \coordinate (A) at (3.5,0.5);
  \coordinate (K) at (3,3.5);
  \coordinate (Kx) at (1.5,3.5);
  \coordinate (Ax) at (3,0.5);
  \coordinate (X) at (5, 0);
  \coordinate (Y) at (0, 5);
  \node[fill=black,circle,inner sep=0.9] (R') at (R) {};
  \node[fill=black,circle,inner sep=0.9] (A') at (A) {};
  \node[fill=black,circle,inner sep=0.9] (K') at (K) {};
  \node[above=8] at (R')                      {\small restaurant};
  \node[above right=-2, align=center] at (A') {\small grocery\\store};
  \node[above right=-2] at (K')               {\small you};
  \draw[<->,line width=0.9] 
    (5,0) -- (O) -- (0,5);
  \draw[vector2,dashed] (K) -- (R) node[midway,left=12,above right=0] {$\vec{fk}$};
  \draw[vector2,dashed] (K) -- (A) node[midway,left=0,above right=0] {$\vec{ka}$};
  \draw[vector] (O) -- (R') node[midway,left=12,above right=0] {$\vec{f}$};
  \draw[vector] (O) -- (K') node[midway,left=12,above right=0] {$\vec{k}$};
  \draw[vector] (O) -- (A') node[midway,left=15,above right=0] {$\vec{a}$};
  \draw[thick] (X)++(0,0.1) --++ (0,-0.2) node[scale=0.9,below=4,left=5] {x};
  \draw[thick] (Y)++(0.1,0) --++ (-0.2,0) node[scale=0.9,below left=1] {y};
\end{tikzpicture}

%% file: figs/fig-coordinates2.tex
\begin{tikzpicture}
  \coordinate (O) at (0,0);
  \coordinate (R) at (1.5,4);
  \coordinate (A) at (3.5,0.5);
  \coordinate (K) at (3,3.5);
  \coordinate (Kx) at (1.5,3.5);
  \coordinate (Ax) at (3,0.5);
  \coordinate (X) at (5, 0);
  \coordinate (Y) at (0, 5);
  \node[fill=black,circle,inner sep=0.9] (R') at (R) {};
  \node[fill=black,circle,inner sep=0.9] (A') at (A) {};
  \node[fill=black,circle,inner sep=0.9] (K') at (K) {};
  \node[above=8] at (R')        {\small The Frogs};
  \node[above right=-2] at (A') {\small Antigone};
  \node[above right=-2] at (K') {\small The Knights};
  \draw[<->,line width=0.9] 
    (5,0) -- (O) -- (0,5);
  \draw[vector2,dashed] (K) -- (R) node[midway,left=12,above right=0] {$\vec{fk}$};
  \draw[vector2,dashed] (K) -- (A) node[midway,left=0,above right=0] {$\vec{ka}$};
  \draw[vector] (O) -- (R') node[midway,left=12,above right=0] {$\vec{f}$};
  \draw[vector] (O) -- (K') node[midway,left=12,above right=0] {$\vec{k}$};
  \draw[vector] (O) -- (A') node[midway,left=15,above right=0] {$\vec{a}$};
  \draw[thick] (X)++(0,0.1) --++ (0,-0.2) node[scale=0.9,below=4,left=5] {\small tragedy};
  \draw[thick] (Y)++(0.1,0) --++ (-0.2,0) node[scale=0.9,below left=1] {\small comedy};
\end{tikzpicture}

%% file: figs/fig-dbs.tex
\begin{tikzpicture}[->,>=stealth',auto,node distance=1.8cm,
  thick,main node/.style={circle,draw,font=\sffamily\Large\bfseries},
  database/.style={
      cylinder,
      shape border rotate=90,
      minimum width = 1.4cm,
      minimum height = 1.0cm,
      aspect=2,
      fill=white,
      draw
    }]
\node (vdb)  [database, align=center, fill=lightgray]                               {};
\node (ve_left)  [process, above of=vdb, xshift=-2.5cm, yshift=1.5cm, align=center] {Vectorization};
\node (ve_right) [process, above of=vdb, xshift=2.5cm, yshift=1.5cm, align=center]  {Vectorization};

\node[draw=black,dotted, fit=(vdb) (ve_left) (ve_right)
     ,inner sep=12mm,label={[xshift=-0.0cm,yshift=-0.2cm,anchor=north]:VDBMS}](FIt1)   {};

\path[->] (vdb) edge [loop above] node {\footnotesize indexing}                    (   );

\node (db1)  [database, left of=ve_left, xshift=-2.5cm, align=center]              {};
\node (db2)  [database, below of=db1, yshift=0.9cm, align=center]                  {};
\node (db2)  [database, below of=db2, yshift=0.9cm, align=center]                  {};

\node (app) [process, right of=ve_right, xshift=3cm, align=center]                 {Application};

\draw [->,out=0,in=0,looseness=0] (db1.east) to node[midway,above,inner sep=5pt,fill=white]{\footnotesize insert}  (ve_left.west);
\draw [->,out=0,in=0,looseness=0] (app.west) to node[midway,above,inner sep=5pt,fill=white]{\footnotesize query}   (ve_right.east);

\draw[->] (ve_left.south)  --++ (0, 0)    |- node[below right,inner sep=5pt]{\footnotesize insert}                    (vdb.west);
\draw[->] (ve_right.south) --++ (0, 0)    |- node[below left,inner sep=5pt] {\footnotesize  query}                    (vdb.east);
\draw[<-] (app.south)      --++ (0, -3.5) -| node[below right,xshift=2.5cm,inner sep=5pt]{\footnotesize return results} (vdb.south);

\node (ii1) [draw=lightgray, fill=mygray, above of=db1, xshift=2.7cm, yshift=1.0cm, align=left] 
                                {  \scriptsize The Frogs: ~~"Shall I crack any of those..."
                               \\\scriptsize The Knights: "Oh! alas! alas! alas!..."
                               \\   \scriptsize Antigone: ~~~"Ismene, dear sister, You..."};
\draw[-, blue] (ve_left.north) edge[out=90,in=270, xshift=-2.5cm, -] (ii1.south);

\node (iq1) [draw=lightgray, fill=mygray, above of=app, xshift=-2.0cm, yshift=1.0cm, align=left] 
                                {\scriptsize "Find the most tragic plays."};
\draw[-, blue] (ve_right.north) edge[out=90,in=270, xshift=2.5cm, -] (iq1.south);

\node (iv1) [draw=lightgray, fill=mygray, below of=db2, xshift=2.2cm, yshift=-2.1cm, align=left] 
                                {  \scriptsize The Frogs: ~~[3, 8, -1, ...]
                               \\\scriptsize The Knights: [6, 7, 0, ...]
                               \\   \scriptsize Antigone: ~~~[7, 1, 5, ...]};
\draw[-, blue] (vdb.south) edge[out=270,in=90, xshift=-1.9cm, yshift=-0.2cm, -] (iv1.north);

\node (iv2) [draw=lightgray, fill=mygray, right of=iv1, xshift=3.2cm, align=left] 
                                {\scriptsize Search vector: [9, -3, 3, ...]};
\draw[-, blue] (vdb.south) edge[out=270,in=90, xshift=2.0cm, yshift=-0.1cm, -] (iv2.north);

\node (ire) [draw=lightgray, fill=mygray, right of=iv2, xshift=2.2cm, align=left] 
                                {\scriptsize 1. Antigone\\
                                 \scriptsize 2. The Knights};
\draw[-, blue] (vdb.south) edge[out=270,in=90, xshift=3.8cm, yshift=-1.0cm, -] (ire.north);

\end{tikzpicture}

%% file: figs/fig-queries.tex
\footnotesize
        \centering
        \hskip0cm\begin{subfigure}[t]{.45\linewidth}
            \begin{tabular}{c}
\begin{lstlisting}[backgroundcolor=\color{light-gray},escapechar=@]
results = collection.search(
    data=[[3, 8, -1, ...]], 
    anns_field="text", 
    param=search_params,
    limit=2,
    expr="country like \"Greece\" && type like \"play\"",
    output_fields=['title'],
    consistency_level="Strong"
)
\end{lstlisting}
             \end{tabular}
             \caption{Query in Milvus}
        \end{subfigure}\hspace{2em}%
        \hskip0cm\begin{subfigure}[t]{.45\linewidth}
            \begin{tabular}{c}
\begin{lstlisting}[backgroundcolor=\color{light-gray}]
payload = {
    "filter": {
        "country": "Greece",
        "type": "play"
    },
    "includeValues": True,
    "includeMetadata": True,
    "topK": 2,
    "vector": [3, 8, -1, ...]
}
response = requests.post(url, json=payload)
\end{lstlisting}
            \end{tabular}
            \caption{Query in Pinecone}
        \end{subfigure}
\par\bigskip 
\hskip0cm\begin{subfigure}[t]{.45\linewidth}
            \begin{tabular}{c}
\begin{lstlisting}[backgroundcolor=\color{light-gray},escapechar=@]
results = collection.query(
    query_embeddings=[[3, 8, -1, ...]],
    n_results=2,
    where={"country": "Greece", "type": "play"}
)
\end{lstlisting}
             \end{tabular}
             \caption{Query in Chroma}
        \end{subfigure}\hspace{2em}%
        \hskip0cm\begin{subfigure}[t]{.45\linewidth}
            \begin{tabular}{c}
\begin{lstlisting}[backgroundcolor=\color{light-gray}]
     SELECT *  
        FROM plays
      WHERE country = 'Greece'
          AND type = 'play'
ORDER BY embedding <-> '[3,8,-1, ...]' 
       LIMIT 2;
\end{lstlisting}
            \end{tabular}
            \caption{Query in PostgreSQL (pgvector)}
        \end{subfigure}

%% file: figs/fig-dbms.tex
\begin{tikzpicture}[->,>=stealth',auto,node distance=1.8cm,
  thick,main node/.style={circle,draw,font=\sffamily\Large\bfseries},
  database/.style={
      cylinder,
      shape border rotate=90,
      minimum width = 1.4cm,
      minimum height = 1.0cm,
      aspect=2,
      fill=white,
      draw
    }]

\node (qc) [process,   xshift=0cm, align=center]                 {Query\\component};
\node (st) [process,   right of=qc, align=center, xshift=1.0cm]  {Storage\\component};
\node (logs) [process, above of=st, align=center]                {Transaction\\logs};
DBMS
\node[draw=black,dotted, fit=(qc) (st) (logs)
    ,inner sep=5mm,label={[xshift=-2.0cm,yshift=-0.2cm,anchor=north]:VDBMS}](FIt1) {};

\node (mem) [process, below of=st, align=center, xshift=-0.0cm, yshift=-0.0cm]  {Memory};
\node (cpu) [process, left of=mem, align=center, xshift=-1.0cm]                 {CPU};
\node (gpu) [process, below of=cpu, align=center]                               {GPU};
\node (oth) [process, below of=mem, align=center]                               {Other\\specialized\\hardware};
\node (db)  [database, right of=mem, align=center, xshift=1.0cm, yshift=-0.0cm] {}; 

\node (software) [startstop, left of=qc, align=center, xshift=-1.2cm]     {Application};

\node (ii2) [draw=lightgray, fill=mygray, right of=db, xshift=2.25cm, yshift=0.66cm, align=left, minimum height = 2.9cm, minimum width = 4.1cm] {};
\node (ii2) [draw=lightgray, fill=mygray, right of=db, xshift=2.1cm, yshift=0.80cm, align=left, minimum height = 2.9cm, minimum width = 4.1cm] {};
\node (ii1) [draw=lightgray, fill=mygray, right of=db, xshift=2.0cm, yshift=1.0cm, align=left] 
                                {\scriptsize id : 18C90122AA...
                               \\\scriptsize Metadata : \{
                               \\\scriptsize ~~type : "play",
                               \\\scriptsize ~~country : "Greece", ...
                               \\\scriptsize \}
                               \\\scriptsize Vector : [7, 1, 5, ...]
                               \\\scriptsize Payload : "Ismene, dear ..."
                               };

\draw [->,out=0,in=0,looseness=0] (software.east) to node[midway,above,inner sep=5pt]{} (qc.west);
\draw [->,out=0,in=0,looseness=0] (qc.east) to node[midway,above,inner sep=5pt]{} (st.west);
\draw [->,out=0,in=0,looseness=0] (st.north) to node[midway,above,inner sep=5pt]{} (logs.south);
\draw [<->,out=0,in=0,looseness=0] (st.south) to node[midway,above,inner sep=5pt]{} (mem.north);
\draw [<->,out=0,in=0,looseness=0] (mem.west) to node[midway,above,inner sep=5pt]{} (cpu.east);
\draw [<->,out=0,in=0,looseness=0] (mem.south) to node[midway,above,inner sep=5pt]{} (oth.north);
\draw [<->,out=0,in=0,looseness=0] (mem.south west) to node[midway,above,inner sep=5pt]{} (gpu.north east);
\draw [<->,out=0,in=0,looseness=0] (mem.east) to node[midway,above,inner sep=5pt]{} (db.west);

\draw[-, blue] (ii1.west) edge[out=180,in=0, xshift=0.0cm, yshift=-0.0cm, -] (db.east);

\end{tikzpicture}

%% file: figs/fig-u-img.tex
\begin{tikzpicture}[->,>=stealth',auto,node distance=1.8cm,
  thick,main node/.style={circle,draw,font=\sffamily\Large\bfseries},
  database/.style={
      cylinder,
      shape border rotate=90,
      minimum width = 1.4cm,
      minimum height = 1.0cm,
      aspect=2,
      fill=white,
      draw
    }]
\node (vdb)  [database, align=center, fill=lightgray]                               {};
\node (ve_left)  [process, above of=vdb, xshift=-2.5cm, yshift=1.0cm, align=center] {Feature\\vectorization};
\node (ve_right) [process, above of=vdb, xshift=2.5cm, yshift=1.0cm, align=center]  {Feature\\vectorization};

\node[draw=black,dotted, fit=(vdb) (ve_left) (ve_right)
     ,inner sep=13mm,label={[xshift=0cm,yshift=-0.2cm,anchor=north]:VDBMS}](FIt1)   {};

\path[->] (vdb) edge [loop above] node {\footnotesize indexing}                    (   );

\node (fea1) [process, left of=ve_left, xshift=-2.5cm,align=center]   {Feature\\extraction};
\node[inner sep=1pt,thick] (bubb) [process, minimum height = 2.3cm, minimum width = 2.3cm, below of=fea1, yshift=-1.9cm, xshift=0.3cm] {};
\node[inner sep=1pt,thick] (buba) [process, minimum height = 2.3cm, minimum width = 2.3cm, below of=fea1, yshift=-1.75cm, xshift=0.15cm] {};
\node[inner sep=0pt,thick] (bub1) [process, below of=fea1, yshift=-1.6cm] {\includegraphics[width=2.3cm]{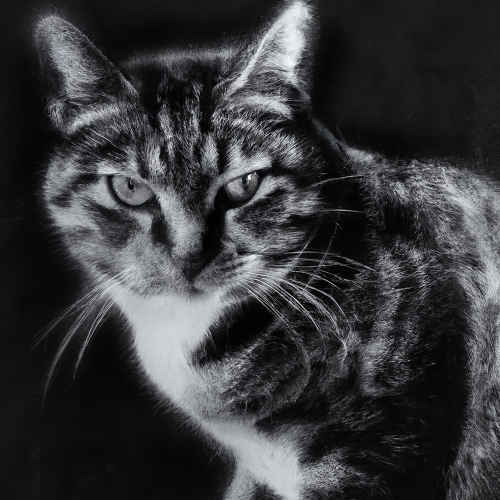}};

\node (fea2) [process, right of=ve_right, xshift=2.5cm, align=center]   {Feature\\extraction};
\node[inner sep=0pt,thick] (bub2) [process, below of=fea2, yshift=-1.6cm]   {\includegraphics[width=2.3cm]{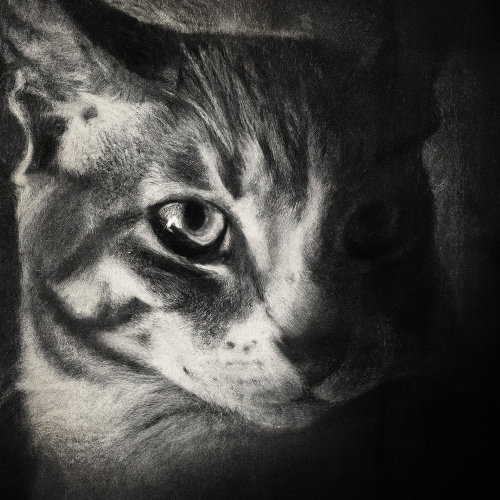}};

\draw [->,out=0,in=0,looseness=0] (bub1.north) to node[midway,above,inner sep=5pt]{\footnotesize }                 (fea1.south);
\draw [->,out=0,in=0,looseness=0] (fea1.east) to node[midway,above,inner sep=5pt,fill=white]{\footnotesize insert} (ve_left.west);

\draw [->,out=0,in=0,looseness=0] (bub2.north) to node[midway,above,inner sep=5pt]{\footnotesize }                (fea2.south);
\draw [->,out=0,in=0,looseness=0] (fea2.west) to node[midway,above,inner sep=5pt,fill=white]{\footnotesize query} (ve_right.east);

\draw[->] (ve_left.south)  --++ (0, 0)    |- node[below right]{\footnotesize insert}                           (vdb.west);
\draw[->] (ve_right.south) --++ (0, 0)    |- node[below left] {\footnotesize query}                            (vdb.east);
\draw[->] (vdb.south)      --++ (0, 0)    |- node[below right,xshift=1cm,inner sep=5pt]{\footnotesize return similar images} (bub2.west);

\end{tikzpicture}

%% file: figs/fig-u-chat.tex
\begin{tikzpicture}[->,>=stealth',auto,node distance=1.8cm,
  thick,main node/.style={circle,draw,font=\sffamily\Large\bfseries},
  database/.style={
      cylinder,
      shape border rotate=90,
      minimum width = 1.4cm,
      minimum height = 1.0cm,
      aspect=2,
      fill=white,
      draw
    }]
\node (vdb-1)   [database, align=center, fill=lightgray]                             {};
\node (ve_left) [process, above of=vdb-1, xshift=-1.5cm, yshift=1.0cm, align=center] {Vectorization};

\node (vdb-2)    [database, right of=vdb-1, xshift=2.0cm, align=center, fill=lightgray] {};
\node (ve_right) [process, above of=vdb-2, xshift=1.5cm, yshift=1.0cm, align=center]    {Vectorization};

\node[draw=black,dotted, fit=(vdb-1) (ve_left)
     ,inner sep=8mm,label={[xshift=0cm,yshift=-0.2cm,anchor=north]:VDBMS (generative model)}](FIt1)   {};

\node[draw=black,dotted, fit=(vdb-2) (ve_right)
     ,inner sep=8mm,label={[xshift=0cm,yshift=-0.2cm,anchor=north]:VDBMS (long-term memory)}](FIt1)   {};

\path[->] (vdb-1) edge [loop above] node {\footnotesize indexing}                  (   );
\path[->] (vdb-2) edge [loop above] node {\footnotesize indexing}                  (   );

\node (db1)  [database, left of=ve_left, xshift=-2.0cm, align=center]              {};
\node (db2)  [database, below of=db1, yshift=0.9cm, align=center]                  {};
\node (db2)  [database, below of=db2, yshift=0.9cm, align=center]                  {};

\node (app) [process, right of=ve_right, xshift=2.5cm, align=center]  {Application};

\draw [->,out=0,in=0,looseness=0] (db1.east) to node[midway,above,inner sep=5pt,fill=white]{\footnotesize insert} (ve_left.west);
\draw [->,out=0,in=0,looseness=0] (app.west) to node[midway,above,inner sep=5pt,fill=white]{\footnotesize query}  (ve_right.east);

\draw [<-,out=0,in=0,looseness=0] (vdb-1.east) to node[midway,above,inner sep=5pt,align=center,fill=white] {\footnotesize query(*)} (vdb-2.west);
\draw [->] (ve_left.south)  --++ (0, 0)    |- node[below left,align=center,xshift=-0.0cm,yshift=1.0cm]    {\footnotesize insert}   (vdb-1.west);

\draw[->] (ve_right.south) --++ (0, 0)    |- node[below right,align=center,yshift=1.0cm] {\footnotesize query \\\footnotesize and insert} (vdb-2.east);
\draw[<-] (app.south)      --++ (0, -4.0) -| node[below right,xshift=3cm,inner sep=5pt]{\footnotesize response} (vdb-1.south);

\draw[dashed,->] (3.8, -1.7) to node[right,yshift=0.25cm]{\footnotesize insert} (vdb-2.south);

\end{tikzpicture}